\documentclass[prb,showpacs,twocolumn,amsmath]{revtex4}
\usepackage{latexsym}
\usepackage{graphicx}

\newcommand{\be}{\begin{equation}}

\newcommand{\ee}{\end{equation}}

\newcommand{\reff}[1]{(\ref{#1})}

\begin{document}

\title{A  Two Rotor Model with spin for  magnetic Nanoparticles }

\author{Keisuke Hatada$^{1,2}$, Kuniko Hayakawa$^{2}$, Augusto Marcelli$^{2}$, Fabrizio Palumbo$^{2}$}

\affiliation{$^1$Scienze e Tecnologie, Universita' di Camerino,
Via Madonna delle Carceri 9, 62032 Camerino, Italy\\
$^2$INFN Laboratori Nazionali di Frascati, 00044 Frascati, Italy}

\begin{abstract}

We  argue that for some species of  magnetic nanoparticles  the macrospin can have a nonvanishing moment of inertia and then an orbital angular momentum. We represent such nanoparticles  by two interacting rigid rotors one of which has a large spin attached to the body, namely    by a Two Rotor Model with spin. By this model we  can describe in a unified way the cases of nanoparticles free and stuck in an  elastic or rigid matrix. We evaluate the magnetic susceptibilities for the latter case and under some realistic  assumptions we get  results in closed form.
 A crossover between thermal and purely quantum hopping occurs at a temperature  much larger than that at which purely quantum tunneling becomes important. A comparison with some experimental data is outlined.

\end{abstract}

\maketitle

1. {\it Introduction}. We consider single domain nanoparticles that can  schematically be represented as a uniform magnetic lattice, the macrospin, rotating in a nonmagnetic lattice.  While it seems natural to  associate   a rigid rotor with the nonmagnetic lattice~[\onlinecite{Chud0}],  the macrospin is generally represented as a pure spin. We think, however, that in some cases, possibly several cases,  also the macrospin should carry a moment of inertia because  the constituting spins will always to some extent drag the orbits.
 One extreme example is given by structures in which   the constituting spins belong to electrons that have such a strong spin-orbit coupling that they are rigidly locked to their orbits. Another example occurs when the macrospin has an electrically charged profile, so that its magnetic moment gets a contribution from the orbital motion. Actually the role of the moment of inertia in the dynamics of a finite system  of particles  was already considered  long  ago for ions in crystals~[\onlinecite{Tram}], and recently for the Scissors Modes of electrons in metal clusters~[\onlinecite{Lipp1}] and quantum dots~[\onlinecite{Serr}] and of ions in crystals~[\onlinecite{Hata, Hata1}]. Concerning nanoparticles,  an inertial parameter was often explicitly introduced in the treatment of tunneling, but only recently, as far as we know, it appeared in the theory of the classical regime~[\onlinecite{Bhat}]. 

 We restrict our attention to  nanoparticles that can be represented as two rigid rotors one of which carries a large spin. For such nanoparticles 
 we adopt a model obtained 
  by a modification of the   Two Rotor Model designed long ago~[\onlinecite{LoIu}] to describe deformed atomic nuclei, in which case the two rotors are the proton and neutron bodies as shown in Fig.1. 
The Two Rotor Model predicts collective excitations called  Scissors Modes,  characterized by a strong  magnetic dipole moment whose coupling with the electromagnetic field  provides their signature. Scissors Modes  have  been  observed for the first time~[\onlinecite{Bohle}] in a rare earth nucleus, $^{156}Gd $, and then  in all deformed atomic nuclei.
 By analogy similar collective excitations were predicted~[\onlinecite{Guer}] and observed~[\onlinecite{Mara}] in Bose-Einstein condensates 
 and predicted (but not yet experimentally searched or found) in several other systems, including 
 metal clusters~[\onlinecite{Lipp}], quantum dots~[\onlinecite{Serr}],
 Fermi~[\onlinecite{Ming}] condensates and crystals~[\onlinecite{Hata, Hata1}].  In all these cases one of the scissors blades must be identified with a structure at rest and the other one with the moving cloud of particles.
    \begin{figure} [here]
  \begin{center}
    \begin{tabular}{cc}
     \includegraphics[width=2cm]{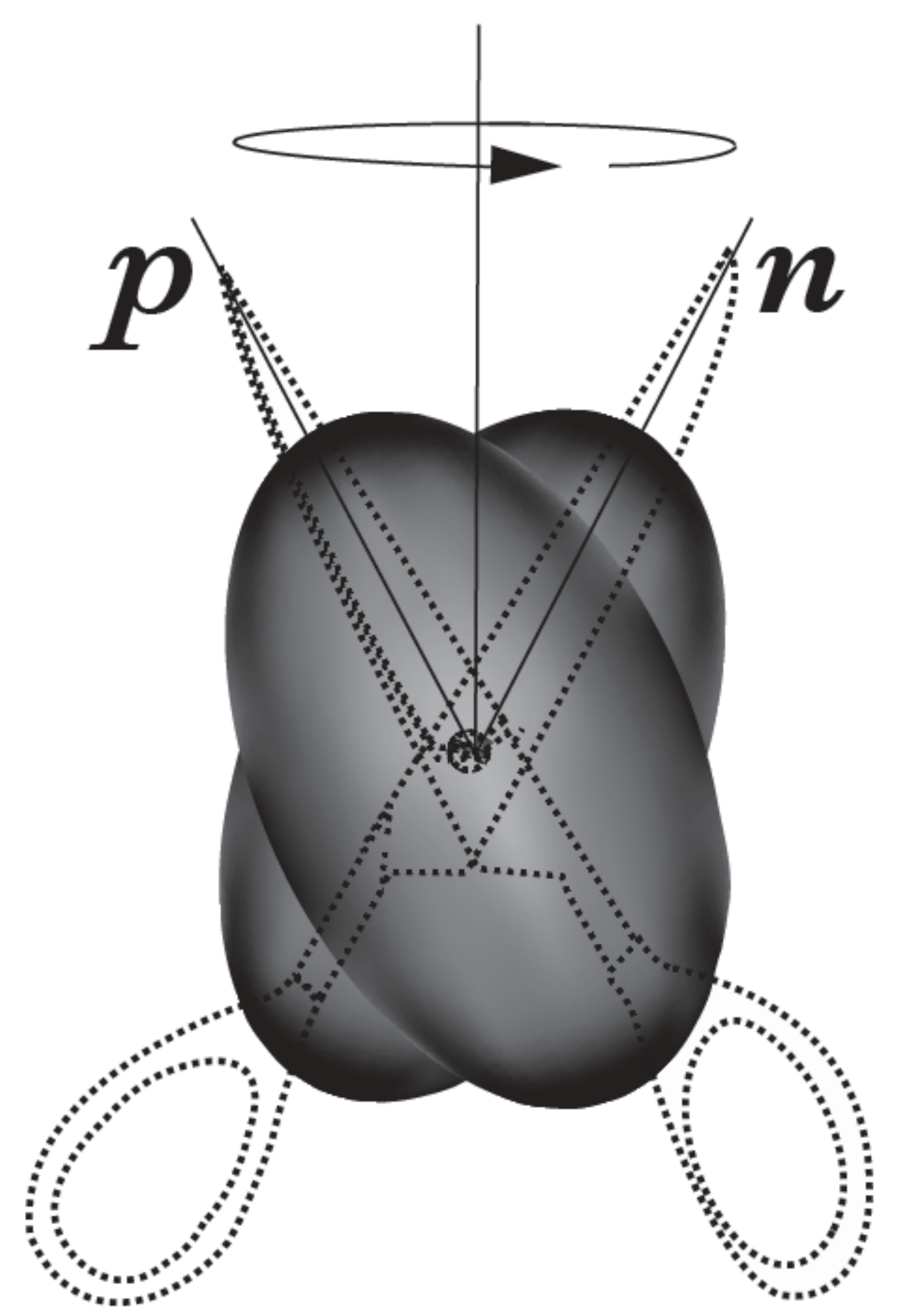}
           \includegraphics[width=2cm]{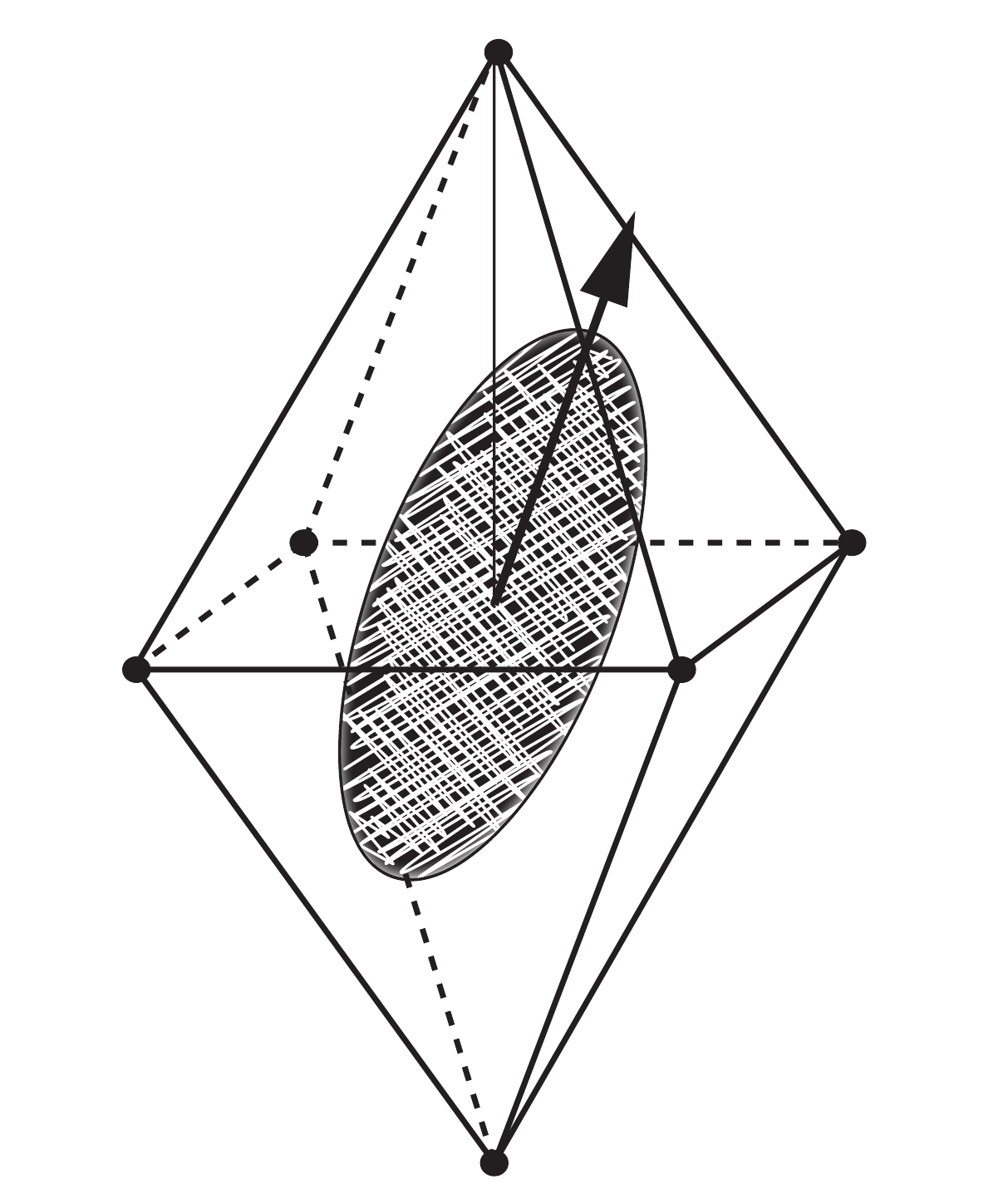}
    \end{tabular}
  \end{center}
 \caption{(a)Scissors Mode in atomic nuclei: the proton (p) and neutron (n) rotors precess around their bisector.
(b)The charge profile of an ion is rigidly locked to its spin, indicated by an arrow.  In both cases the lowest excited states are called Scissors Modes.  }
\end{figure}

In order to put our model into  perspective  let us examine the characteristic features of the magnetic susceptibilities of nanoparticles. 
 If the nanoparticles are free or  stuck in an elastic matrix, one must take into account  effects due to angular momentum conservation, because rotation of the macrospin entails rotation of the nonmagnetic lattice~[\onlinecite{Chud2}].
If instead they are  stuck in a rigid matrix, one can study the rotation of the macrospin in the frame of reference of the nonmagnetic lattice. In this case the magnetic  susceptibilities  are characterized by a crossover at a low temperature $T_c$. Their theoretical description above $T_c$ can be obtained by classical thermodynamics or the Landau-Lifshitz-Gilbert equation~[\onlinecite{LLG}] that describes the precession of a magnetic moment under the effect of different forces. Above $T_c$, therefore, magnetic susceptibilities show an essentially classical behaviour.  Below $T_c$, on the contrary, tunneling was advocated  to explain the puzzling behaviour of Weil's nickel particles~[\onlinecite{Weil, Bean, Chud}]. The  crossover therefore  has been  explained as a change from thermal hopping to quantum tunneling~[\onlinecite{Grab}]. Its nature has been further studied in Refs.~[\onlinecite{Chud1, Chud2}].

All these situations can be studied in a unified framework by a Two Rotor Model with spin. Therefore, having in mind future applications, we will present the model in its full generality, even though detailed calculations have been  performed only for nanoparticles stuck in a rigid mathrix.

Before proceeding, we want to stress  an analogy between a nanoparticle stuck in a rigid matrix and a magnetic ion  in a crystal cell. Such an analogy is closer 
 when the electrons of the ions that carry the magnetism have such a strong spin-orbit coupling that their charge density profile is rigidly locked  with the spin, as shown in Fig.{1}.  In a recent work~[\onlinecite{Hata1}] we adapted the Two Rotor Model to such a system representing the magnetic ion with spin-orbit locking as a rigid rotor with spin. In the present paper we will exploit this analogy to study in detail nanoparticles stuck in a rigid matrix. We will restrict ourselves to axially symmetric nanoparticles, so that  the component of the orbital angular momentum along the symmetry axes of the rotors vanishes.

2. {\it The Two Rotor Model with spin}. The general form of the Two Rotor Model Hamiltonian~[\onlinecite{LoIu}] is 
\be
H =\frac{1}{2{\mathcal I}_1}  {\vec L}_1^2 +  \frac{1}{2{\mathcal I}_2} {\vec L}_2^2 +V
\ee
where ${\vec L}_1, {\vec L}_2  $,  ${\mathcal I}_1, {\mathcal I}_2$  are the angular momenta and moments of inertia of  the nonmagnetic lattice and macrospin respectively,  and $V$  the sum of the potential interaction between them plus an external potential.  
  We  denote by  ${\hat \zeta}_1, {\hat \zeta}_2$ the symmetry axes of the nonmagnetic lattice and the macrospin,  and  assume   the easy axis of magnetization along ${\hat \zeta}_1$ and the spin  along $ {\hat \zeta}_2$.
 We write the wave functions of the macrospin as functions of the polar angles, $\psi=\psi(\theta_2, \phi_2)$, with the understanding that the spin has the direction of $\vec \zeta_2$. We can invert the direction of the spin by performing an inversion of  $\vec \zeta_2$. We can  construct even and odd wave functions with respect to spin inversion $ I_s$: 
$
I_s \psi_{\sigma}(\theta, \varphi) = \sigma \psi_{\sigma}(\theta, \varphi)\,,  \,\,\,\sigma=\pm1\,.
$
 The interaction between the two rotors will depend only  on the angle $2 \theta$ between  ${\hat \zeta}_1, {\hat \zeta}_2$, 
$
\cos(2\theta)= {\hat \zeta}_1 \cdot {\hat \zeta}_2\,.
$
The independent variables are ${\hat \zeta}_1, {\hat \zeta}_2 $.  They can be replaced  by variables that describe the system of the rotors as a whole plus the variable $\theta$.  
To this end we define a frame  of axes
\be
{\hat \xi}=\frac{{\hat \zeta}_2 \times {\hat \zeta}_1}{ 2 \sin\theta}, \, 
{\hat \eta}=\frac{{\hat \zeta}_2 - {\hat \zeta}_1}{2 \sin\theta}, \,
{\hat \zeta}=\frac{{\hat \zeta}_2 + {\hat \zeta}_1}{2 \cos \theta}
\ee
and denote by $\alpha, \beta, \gamma$ its Euler angles. 
The correspondence $\{ {\hat \zeta}_1, {\hat \zeta}_2 \} = \{ \alpha, \beta, \gamma, \theta \}$ is one-to-one and regular for $ 0< \theta < \frac{\pi}{2} $. The variables $\{ {\hat \zeta}_1, {\hat \zeta}_2 \} = \{ \alpha, \beta, \gamma, \theta \}$
are not sufficient to describe the configurations of the classical system, but they describe uniquely the quantized system owing to the constraints
$
{\vec L}_1 \cdot {\hat \zeta}_1= 0, \,\,\, {\vec L}_2 \cdot {\hat \zeta}_2= 0 \label{constraints}
$
necessary for rigid bodies with axial symmetry. These constraints are automatically satisfied if we take the wave functions to depend on ${\hat \zeta}_1, {\hat \zeta}_2 $ only. Because of these constraints the  component of the total angular momentum of the macrospin along $\vec \zeta_2$ is constant and equal to the spin.

We can now transform the Hamiltonian in the new variables. We define  the operators
 \be
{\vec L}={\vec L}_1 + {\vec L}_2, \,\,\, 
{\vec {\mathcal L}}= {\vec L}_1 - {\vec L}_2 \,.
\ee
${\vec L}$ is the total orbital angular momentum acting on the Euler angles $\alpha, \beta, \gamma $, while ${\vec {\mathcal L}}$ is not an angular momentum, and has the  representation~[\onlinecite{LoIu}]
\be
{\vec {\mathcal L}}_{\xi}= i \frac{\partial}{\partial \theta}, \,\,\,  {\vec {\mathcal L}}_{\eta}= -\cot \theta {\vec L}_{\zeta}, \,\,\,
 {\vec {\mathcal L}}_{\zeta}=-\tan \theta  {\vec L}_{\eta} \,.
\ee 
The transformed  Hamiltonian  is the sum of the rotational kinetic energy of the two rotors system as a whole plus an intrinsic energy
\be
H= \frac{{\vec {\mathcal L}}^2 }{2{\mathcal I}}+ H_I 
\ee
where
$
{\mathcal I}= {\mathcal I}_1 {\mathcal I}_2/
({\mathcal I}_1 + {\mathcal I}_2)\,.
$
The intrinsic energy reads
\begin{eqnarray}
H_I  = \frac{1}{2{\mathcal I}}\left[ \cot^2\theta {\mathcal L}_{\zeta}^2 + \tan \theta^2 {\mathcal L}_{\eta}^2 - \frac{\partial^2}{\partial \theta^2}
-2 \cot(2\theta) \frac{\partial}{\partial \theta}\right] &&
\nonumber\\
+ \frac {{\mathcal I}_1 - {\mathcal I}_2}{4 {\mathcal I}_1 {\mathcal I}_2} 
\left[ -  \tan \theta {\mathcal L}_{\zeta} {\mathcal L}_{\eta} -  \cot\theta {\mathcal L}_{\eta}  {\mathcal L}_{\zeta}      
+ i  {\mathcal L}_{\zeta} \frac{\partial}{\partial \theta} \right] +V \,.&&
\end{eqnarray}
This Hamiltonian was studied in detail~[\onlinecite{LoIu}] for ${\mathcal I}_1 \sim {\mathcal I}_2 $ and small $\theta$, as appropriate to atomic nuclei.

 If the nanoparticle is stuck  in a rigid matrix, the Hamiltonian represents the energy of the macrospin in the frame of reference of the nonmagnetic lattice
\be
H =   \frac{\hbar^2}{2{ \mathcal I}} \left( - \frac{\partial^2}{\partial \theta^2} - \cot \theta  \frac{\partial}{\partial \theta}-
\frac{1}{\sin^2\theta} \frac{\partial^2}{\partial \phi^2} \right) +V   \label{Hoscill}
\ee
where now $\theta$, not $2 \theta$,  is the angle between the $\zeta_2, \zeta_1$-axes and ${\mathcal I}$ the moment of inertia   with respect to the $\xi$- and $\eta$-axes. In the presence of an ac magnetic field ${\vec H}_{ac}$ of angular frequency $\omega$ we assume the potential 
\be
 V = \frac{1}{2}C \sin^2 \theta - {\vec \mu} \cdot {\vec H}_{ac}\cos (\omega t) \,,
\ee
where ${\vec \mu}$ is the  magnetic moment of the macrospin  and $C$ a restoring force constant due to the electric  and  magnetic interactions of the macrospin with the nonmagnetic  lattice.
In the present preliminary investigation we consider one noninteracting nanoparticle in an ac magnetic field. We assume that the macrospin does not have an electrically charged profile, so that there is no orbital contribution to the magnetic moment.

3. {\it Reactive susceptibility and spectrum of the Two Rotors Model}.
We report only the study of the reactive susceptibility whose general expression  for one nanoparticle is
\be
\chi' = \frac{1}{Z}\sum_{\alpha \alpha'}<\alpha| \cos \theta |\alpha'><\alpha'| \vec{\mu}\cdot \vec{H}_{ac}|\alpha> \Phi_{\alpha, \alpha'}
\ee
where
\be
\Phi_{\alpha, \alpha'}  =  \left(\nu_{\alpha}- \nu_{\alpha'} \right)\frac{E_{\alpha'}-E_{\alpha}-\hbar \omega}{ \left( E_{\alpha'}-E_{\alpha}-\hbar \omega \right)^2 + \Gamma^2} \,.      \label{suscept}
\ee
In the last equation 
$
\nu_{\alpha}= \exp \left( - E_{\alpha} / k_BT \right),  \,\,\, Z= \sum_{\alpha} \nu_{\alpha}, \, k_B$ being the Boltzman constant.
  $\Gamma$ is the amplitude of spontaneous transitions between the levels $\alpha, \alpha'$ assumed to be  level independent. We neglected the elastic term that contributes only for $\omega=0$.

The representation  of  nanoparticles by rigid rotors, however, cannot be valid at all temperatures. There will be a maximum temperature $T_M$, beyond which the nanoparticle will change shape or will be altogether demagnetized. This will provide an effective  cut off to the spectrum,
that we divide  into three regions.

{\it Region I, $E_{\alpha} << C$, harmonic oscillator modes}. For the lowest lying states the potential can be approximated by a double well.
For $0< \theta < \pi/2$   the Hamiltonian  becomes
\be
  H=  \,\frac{\hbar^2}{2\mathcal I}\,\biggl(\,- \frac{\partial^2}{\partial\,\theta^2}
  - \frac{1}{\theta}\,\frac{\partial}{\partial\,\theta}
  - \frac{1}{\theta^2}\, \frac{\partial^2}{\partial \phi^2}  \,\biggr) + \frac{1}{2} \,C\,\theta^2 \,.
  \ee
 We recognize  the Hamiltonian of a two dimensional harmonic oscillator if we identify $\theta$ with the polar radius. We denote by  $\psi_{nm}(\theta, \phi)$ the  eigenfunctions of the harmonic oscillator normalized  in  $0< \theta<\infty$, for which we adopt the notation of Ref. [\onlinecite{Hata}].
 We obtain the Hamiltonian and the eigenfunctions
for $\pi/2 < \theta < \pi $ by the change $\theta \rightarrow \pi - \theta $.
The (nonnormalized) eigenfunctions of definite symmetry with respect to spin inversion in the entire range $0< \theta < \pi$ are
  \begin{eqnarray}
\psi_{nm\sigma}(\theta, \phi)&=&\frac{1}{\sqrt 2}\left[ \psi_{nm}(\theta, \phi)+ \sigma \,\psi_{nm}(\pi- \theta, \pi + \phi) \right]
\nonumber\\\
&+& \delta \psi_{nm\sigma}(\theta, \phi)\,, \,\,\, \sigma=\pm1
\end{eqnarray}
where $\delta \psi_{nm\sigma}(\theta, \phi) $ is the distortion due to tunneling.
 They have  eigenvalues 
 \be
 E_{nm\sigma}= \hbar \Omega \,(\,2n + |m| +1\,)- \frac{1}{2} \sigma \delta  E_{nm}
 \ee
 where
 $
  \Omega = \sqrt {C /{\mathcal I}}= C \theta_0^2/  {\hbar}, \,\,\,  
\theta_0^2=
  \hbar /  \sqrt{{\mathcal I} C}  \label{Omega}
 $
and $\delta  E_{nm} $ is the energy splitting  due to tunneling.
  We quote the values of $\theta_0$ in other systems:   $\theta_0^2 \le  0.5$  in  $LaMnO_3$ [\onlinecite{Hata,Hata1}] while   in the atomic nuclei of  the rare earths [\onlinecite{LoIu}]  $\theta_0^2\approx 10^{-2}$.

The  energy splitting of  the ground state is very  small,
$
\delta E \sim \hbar \Omega \exp \left(-2/ \theta_0^2\right) \label{splitting}.
$
 Nevertheless for sufficiently low temperatures tunneling transitions will become important. This will happen when the tunneling temperature 
\be
 T_t = \frac{\hbar \Omega} {k_B}  \exp \left(-2/ \theta_0^2 \right) \label{T_t}
\ee
is reached.

{\it Region II, $E_{\alpha}>>C$, free rotor modes}. At  energies much higher than the potential barrier, the nanoparticle  will have the spectrum of a free rotor
\be
E_l=  \frac{\hbar^2}{2{ \mathcal I}}l(l+1) \label{freerotor}   \,.
\ee
Here the wave functions are the spherical functions whose spin-parity is
$
I_s \psi_{lm}(\theta, \phi)= (-)^l  \psi_{lm}(\theta, \phi)\,.
$

{\it Region III, $E_{\alpha} \sim C$}. In this region for $E_{\alpha}< C$ the states are characterized by tunneling, whose amplitude is no longer negligible, while the energy splitting can become comparable to the energy $\hbar \omega$. As a consequence the strongest dependence on the ac frequency should come from this part of the spectrum. Notice that  we expect in our model two types of tunneling effects: those arising from region III of the spectrum at a temperature close to the blocking temperature, and those appearing   at  a temperature much smaller than the crossover temperature.
Both types of effects give rise to a considerable complexity in the calculations. We avoid the tunneling effects at low temperature restricting ourselves to $T>T_t$, and those at higher temperature   eliminating region III   of the spectrum by extending the regions I and II  to an energy $k_B T_*$ such that 
\begin{eqnarray}
  E &=& 
  \left\{ 
    \begin{array}{rl}
      & E_{nm} \,,\hspace{6mm}E_{nm} < k_BT_*\\
      & E_l\,,\hspace{5mm} k_BT_* < E_l <  k_BT_M\,.
    \end{array}
     \right.  \label{T_*}
\end{eqnarray}
In this way renounce to derive the dependence of the susceptibility on the ac frequency, and we are forced to assume that  $\hbar \omega
<<|E_{\alpha} - E_{\alpha}'| $ in the entire spectrum. We note that under such condition the function $\Phi_{\alpha, \alpha'}$ appearing in the expression of the susceptibility becomes symmetric.

 The temperature $T_*$ that optimizes our approximation can in principle be evaluated, but for the sake of simplicity we   will assume it  as a phenomenological parameter.

4. {\it Evaluation of the reactive  susceptibilitiy}. We evaluate the longitudinal susceptibility, namely the response of the component of the magnetic moment along the ac magnetic field. Its average over the directions of the easy axis of the nanoparticle, namely over the directions of $\hat \zeta_1$ gives
\begin{eqnarray}
\chi' (x) =  \frac{4}{3} \chi_T'(x) + \frac{2}{3}  \, \chi_L'(x)  
\end{eqnarray}
where $ \chi_T'(x),  \chi_L'(x)   $ are the contributions arising when  $\hat \zeta_1$ is respectively tranverse, parallel to the ac magnetic field
\begin{eqnarray}
 \chi_T'(x) &=& \frac{1}{Z}\sum_{\alpha \alpha'} <\alpha|\sin \theta \cos \phi \, |\alpha'>^2 \Phi_{\alpha, \alpha'}
 \nonumber\\
  \chi_L'(x) &=&\frac{1}{Z}\sum_{\alpha \alpha'}<\alpha| \cos \theta \, |\alpha'>^2 \Phi_{\alpha, \alpha'} \,.
\end{eqnarray}
 The contribution of the free rotor modes turns out to be very small for reasonable values of the parameters, and therefore we neglect it in this preliminary report. Therefore  the quantum number  $\alpha$   represents the oscillator quantum numbers $(n,m)$, and the  sums extend in the range given by Eq.\reff{T_*}, with $T_*=T_M$. 
 It is convenient to introduce  the parameters
 \be
  \rho_C= \frac{C}{k_BT_*}\,, \,\, \, \rho_{\Omega}=  \frac{\hbar \Omega}{k_B T_*}= \rho_C \theta_0^2
   \,,      \label{parameters}
\ee
  the reduced temperature
 \be
 x=\frac{T}{T_*}
 \ee
 and the constant 
 \be
\chi_0=  \mu \,  \frac{\mu H_{ac}}{k_B T_*}  \, \frac{1}{2 \rho_C^2} \,.
\ee
After evaluation of the matrix elements to lowest order in $\theta_0$ we get for  the longitudinal part 
  \begin{eqnarray}
 \chi_L'(x)&= &\frac{1}{ Z} \frac{1}{2} \rho_{\Omega} \chi_0 \left[ 1-
 \exp\left(- \frac { \rho_{\Omega}}{x}  \right)\right] \sum_{nm}      (n+1)
 \nonumber\\
 && \times  (n+|m|+1)  \exp\left(- \frac {\rho_{\Omega}}{x}(2n+|m|)  \right)  \label{exact}
 \end{eqnarray}
 with a similar but longer expression for $\chi_T'$  that will be reported somewerelse. The sums can be performed exactly using  the generating function
 \be
 Z_{\rho_1,\rho_2} =  \sum_{nm}  \exp\left(-2 \rho_1 n - \rho_2m  \right) 
 \ee
and  the identity
\begin{eqnarray}
&&  \sum_{nm} n^h m^k \exp\left(- \rho (2 n+m ) \right) 
 \nonumber\\ 
  && = (-2)^{-h} (-1)^k\frac{\partial^{h+k}}{\partial \rho_1^h \partial \rho_2^k} Z_{\rho_1,\rho_2}|_{\rho_1=\rho_2=\rho}\,.
\end{eqnarray} 
The partition function is given by
$
Z= Z{\rho_1=\rho_2 =  \rho_{\Omega}/x}\,.
$
The resulting expressions, however are rather lengthy. They simplify  for 
  \be
 \rho_{\Omega} / x  << 1 \,, \label{condition}
\ee
in which case we can approximate the sums by integrals, getting
 \be
 \chi_T'(x) = \chi_0 \,\frac{8}{3}   \rho_C f_T(x)\,, \,\,\, \chi_L'(x) =\chi_0  f_L(x) \label{approx}
\ee
where
\be
f_T(x) =g(x)^{-1}
\left[1 - \left( 1+ \frac{1}{x}   + \frac{1}{2x^2}  \right) \exp \left( -\frac{1}{x} \right)\right] 
\ee
\begin{eqnarray}
f_L(x)&=& g(x)^{-1} x \left[1 - \left( 1+ \frac{1}{x}   + \frac{1}{2x^2} +  \frac{1}{6x^3} \right) \right.
\nonumber\\
&&  \left. \times  \exp \left( -\frac{1}{x} \right) \right]
\end{eqnarray}
with
\be
g(x)= 1-\left(1+\frac{1}{x} \right) \exp \left( -\frac{1}{x} \right) \,.
\ee
The functions $\chi_L(x)'/\chi_0, \chi_(x)'/\chi_0, \chi_(x)_T'/\chi_0$ are plotted in Fig.2 for $\rho_C= 0.015$, chosen to have a typical shape of the nanoparticle susceptibility. Because $ \rho_{\Omega}< \rho_C$, such a value justifies the approximation of sums by integrals.

In the continuous approximation we get a classical scaling because $\chi(x)'/\chi_0$ depends only on the classical parameter $\rho_C$.
According to the condition \reff{condition}, however, such approximation will only be valid  for temperatures
 \be
T> T_c = \frac{\hbar \Omega}{k_B} =\rho_{\Omega} T_* \,. \label{rho}
\ee
For $T_t < T< T_c$ we must use the exact expressions \reff{exact}. We observe that according to Eq.~\reff{T_t} $T_t \approx T_c \exp(-2/ \theta_0^2 )$, and since $\theta_0^2$ must be  smaller than 1 for the macrospin to be polarized, we conclude that $T_t << T_c$. Because of the smallness of $T_t$ and because $\chi_L'$ and $ \chi_T' $  are smooth functions, we can  assume that  
\be
\chi_L' (T_t) \approx \chi_L' (0)= \frac{1}{2}\rho_{\Omega} \chi_0\,,  \,\,\,\chi_T' (T_t)\approx \chi_T' (0)= \frac{8}{3}\rho_C \chi_0
\ee
where $ \chi_L' (0), \chi_T' (0) $ are the exact values neglecting   tunneling.
 We see that the continuous approximation is valid for  $\chi_T' $ up to small temperatures, but  instead the exact expression of $\chi_L' $, at variance with \reff{approx},  
 does depend on $\rho_{\Omega}$, a quantum parameter, and does not go to zero. From the plot of the exact expression of  $\chi_L' $ (not reported here for lack of space) we see that  it has an inflection at a value of the reduced temperature  that decreases with  decreasing $\rho_{\Omega}$.
Because  above  $T_c$ the energy can be approximated by a continuous variable, while below the quantum spectrum must be used,   $T_c$ is the temperature of crossover from classical to quantum behavior.

\begin{figure}
      \begin{tabular}{ll}
                 \includegraphics[width=7cm]{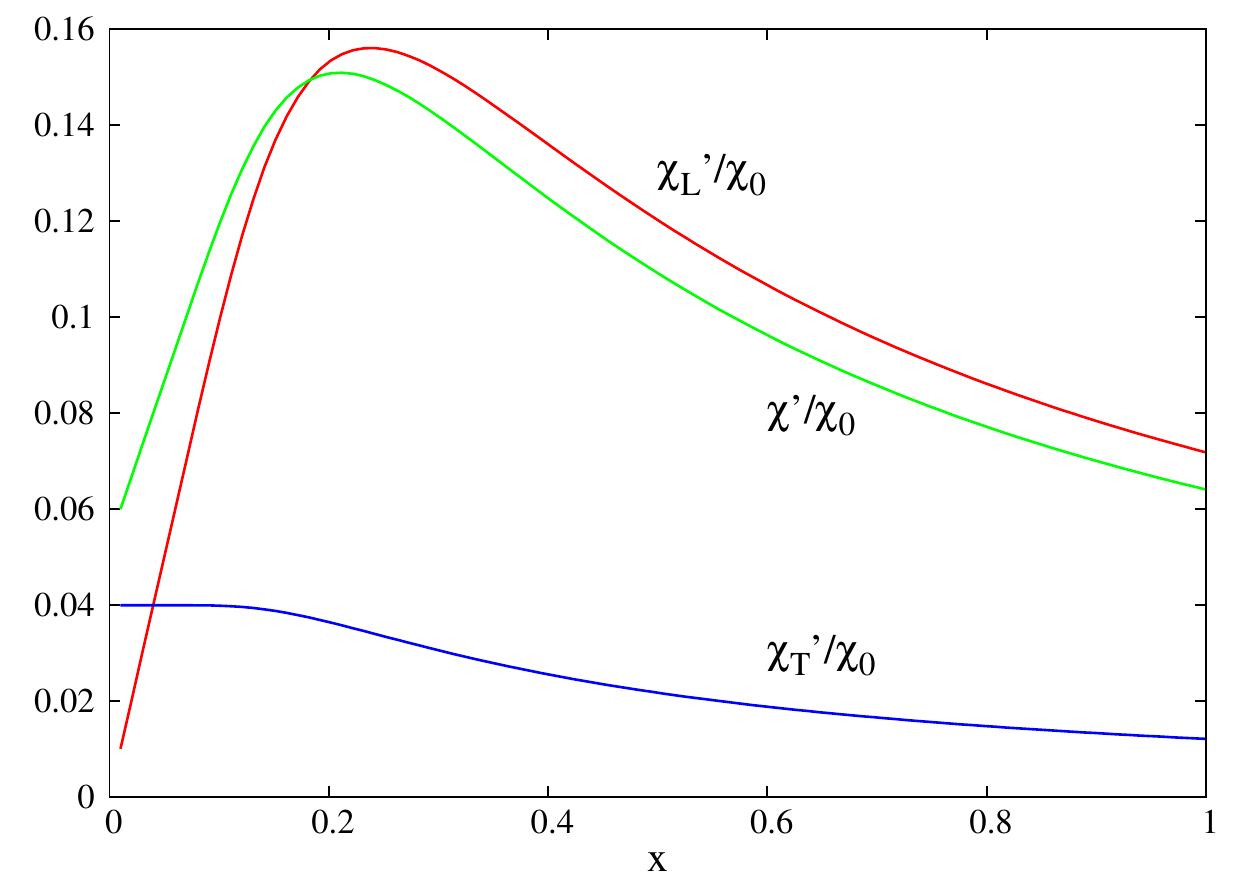}

    \end{tabular}
 \caption{The functions  $ \chi_L '/ \chi_0$ (red line, upper)$, \chi '/ \chi_0$ (green line, middle) and  $\chi_T '/ \chi_0$ (blue line, lower)  versus the reduced temperature $x$  for $\rho_C=0.015$.}
     \label{f1}
\end{figure}

 5. {\it How to compare with experiment}. The susceptibility depends on the volume $v$ of the nanoparticles. In our model such  dependence  is contained in the parameters  ${\mathcal I}, C$ and $T_*$. For $T>T_c$, however, the function $\chi' / \chi_0$ depends only on $\rho_C$, and it should not be affected significantly by a variation of this parameter around its small value. Moreover  $\rho_C= C/ k_B T_* $ should not depend on the volume.  
  Indeed $T_*$ is approximately the temperature at which the macrospin becomes a free rotor, and therefore it should be proportional to the potential barrier $C$, that in turn should be proportional to $v$. Therefore  in the scaling region the average over the volume distribution does not alter the shape of $\chi' / \chi_0$:  All experimental curves, once plotted as functions of the reduced temperature, should  become identical with one another, even though  such universality can to some extent be due to the replacement of region III of the spectrum by the extension of regions I and II. For $T>T_c$, therefore
   
1) the parameter $\rho_C$  can be determined by  fitting the experimental susceptibility 
 
2)  the temperature $ T_* = T/x $ can be evaluated  for $x=x_B$, the reduced temperature at which $\chi'$ has a maximum,  determined theoretically, and  the corresponding experimental value  $T_B$.

To make a complete  comparison  we must perform  a numerical average over the exact expression of the susceptibility. In such a case the comparison with data would fix also $ \rho_{\Omega}$ and we could evaluate 
 the crossover temperature  $T_c$  from its definition $T_c = \rho_{\Omega}T_*$ and compare with the value $x_c$ at which $\chi'$ has an inflection, and estimate  the tunneling temperature $T_t = \rho_{\Omega} T_* \exp ( -2 \rho_C/ \rho_{\Omega})$.
 
  We quote a number of examples~[\onlinecite{Examples}], without attempt to completeness,  in which  the  susceptibility has a form  that appears to us compatible with  our model.  In these experiments the blocking temperature is at most of the order of  one hundred degrees, so that 
  $T_*$ is  a few hundreds  degrees  or smaller and according to Eq.\reff{rho} $  T_c $ is of the order of 1 degree.
 There are also cases in which the form of the reactive susceptibility is not compatible with our model. We quote 2 examples. In the first one~[\onlinecite{Bomb}] there are 2  magnetic atoms, namely TM and RE,  in the second~[\onlinecite{Taji}]  different phases. No wonder that our model, in which the macrospin is assumed to have a uniform structure, cannot reproduce these data.

\appendix

\subsection *{Acknowledgment}
We are grateful to D. Di Gioacchino for interesting discussions.

\end{document}